\documentclass[sigconf]{acmart}
\AtBeginDocument{%
  }
\def\refname{REFERENCES}

\setcopyright{acmlicensed}
\copyrightyear{2018}
\acmYear{2018}
\acmDOI{XXXXXXX.XXXXXXX}
\acmConference[Conference acronym 'XX]{Make sure to enter the correct
  conference title from your rights confirmation email}{June 03--05,
  2018}{Woodstock, NY}
\acmISBN{978-1-4503-XXXX-X/2018/06}

\usepackage{multirow} 
\usepackage{enumitem} 
\usepackage{subfigure}
\copyrightyear{2025}
\acmYear{2025}
\setcopyright{cc}
\setcctype{by}
\acmConference[SIGIR '25]{Proceedings of the 48th International ACM SIGIR Conference on Research and Development in Information Retrieval}{July 13--18, 2025}{Padua, Italy}
\acmBooktitle{Proceedings of the 48th International ACM SIGIR Conference on Research and Development in Information Retrieval (SIGIR '25), July 13--18, 2025, Padua, Italy}\acmDOI{10.1145/3726302.3729881}
\acmISBN{979-8-4007-1592-1/2025/07}

\patchcmd{\abstractname}{Abstract}{\MakeUppercase{Abstract}}{}{}

\begin{document}

%%
%% The "title" command has an optional parameter,
%% allowing the author to define a "short title" to be used in page headers.
\title{A Learnable Fully Interacted Two-Tower Model for Pre-Ranking System}

%%
%% The "author" command and its associated commands are used to define
%% the authors and their affiliations.
%% Of note is the shared affiliation of the first two authors, and the
%% "authornote" and "authornotemark" commands
%% used to denote shared contribution to the research.
\author{Chao Xiong}
\authornote{Equal Contributions.}
%%\authornote{Both authors contributed equally to this research.}
\orcid{0007-0372-9277}
\affiliation{%
  \institution{Ant Group}
  \city{Hangzhou}
  \country{China}
}
\email{xc272640@antgroup.com}

\author{Xianwen Yu}
\orcid{0005-6402-6642}
\authornotemark[1]
\authornote{Corresponding author.}
\affiliation{%
  \institution{Ant Group}
  \city{Hangzhou}
  \country{China}
}
\email{yuxianwen.yxw@antgroup.com}

\author{Wei Xu}
\orcid{0007-4709-260X}
\affiliation{%
  \institution{Ant Group}
  \city{Hangzhou}
  \country{China}
}
\email{weicheng.xw@antgroup.com}

\author{Lei Cheng}
\orcid{0002-2186-699X}
\affiliation{%
  \institution{Ant Group}
  \city{Hangzhou}
  \country{China}
}
\email{lei.chenglei@antgroup.com}

\author{Chuan Yuan}
\orcid{0000-9018-8374}
\affiliation{%
  \institution{Ant Group}
  \city{Hangzhou}
  \country{China}
}
\email{yuanzheng.xy@antgroup.com}

\author{Linjian Mo}
\orcid{0002-6682-1448}
\affiliation{%
  \institution{Ant Group}
  \city{Hangzhou}
  \country{China}
}
\email{linyi01@antgroup.com}

%%
%% By default, the full list of authors will be used in the page
%% headers. Often, this list is too long, and will overlap
%% other information printed in the page headers. This command allows
%% the author to define a more concise list
%% of authors' names for this purpose.
\renewcommand{\shortauthors}{Chao Xiong et al.}

%%
%% The abstract is a short summary of the work to be presented in the
%% article.
\begin{abstract}
Pre-ranking plays a crucial role in large-scale recommender systems by significantly improving the efficiency and scalability within the constraints of providing high-quality candidate sets in real time. The two-tower model is widely used in pre-ranking systems due to a good balance between efficiency and effectiveness with decoupled architecture, which independently processes user and item inputs before calculating their interaction (e.g. dot product or similarity measure). However, this independence also leads to the lack of information interaction between the two towers, resulting in less effectiveness. In this paper, a novel architecture named \textit{learnable Fully Interacted Two-tower Model} (FIT) is proposed, which enables rich information interactions while ensuring inference efficiency. FIT mainly consists of two parts: Meta Query Module (MQM) and Lightweight Similarity Scorer (LSS). Specifically, MQM introduces a learnable item meta matrix to achieve expressive early interaction between user and item features. Moreover, LSS is designed to further obtain effective late interaction between the user and item towers. Finally, experimental results on several public datasets show that our proposed FIT significantly outperforms the state-of-the-art baseline pre-ranking models.
\end{abstract}

%%
%% The code below is generated by the tool at http://dl.acm.org/ccs.cfm.
%% Please copy and paste the code instead of the example below.
%%
\begin{CCSXML}
<ccs2012>
   <concept>
       <concept_id>10002951.10003317</concept_id>
       <concept_desc>Information systems~Information retrieval</concept_desc>
       <concept_significance>500</concept_significance>
       </concept>
   <concept>
       <concept_id>10010147.10010257</concept_id>
       <concept_desc>Computing methodologies~Machine learning</concept_desc>
       <concept_significance>500</concept_significance>
       </concept>
 </ccs2012>
\end{CCSXML}

\ccsdesc[500]{Information systems~Information retrieval}
\ccsdesc[500]{Computing methodologies~Machine learning}

%%
%% Keywords. The author(s) should pick words that accurately describe
%% the work being presented. Separate the keywords with commas.
\keywords{Recommender Systems,Pre-Ranking System,Neural Networks}
%% A "teaser" image appears between the author and affiliation
%% information and the body of the document, and typically spans the
%% page.
%%
%% This command processes the author and affiliation and title
%% information and builds the first part of the formatted document.
\maketitle

\section{INTRODUCTION}
\begin{figure}[t]
    \centering
    \subfigure[Cascade Ranking System] {
        \includegraphics[width=0.35\linewidth]{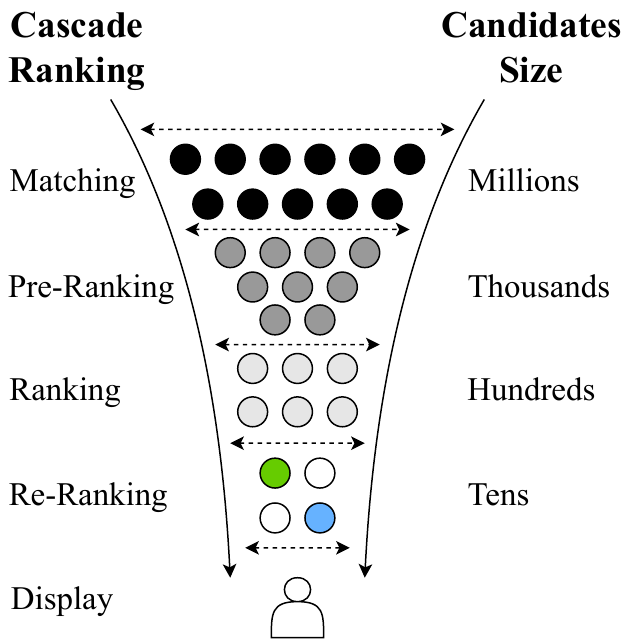}
        \label{fig:1a}
    }
    \subfigure[Two-Tower Model] {
        \includegraphics[width=0.55\linewidth]{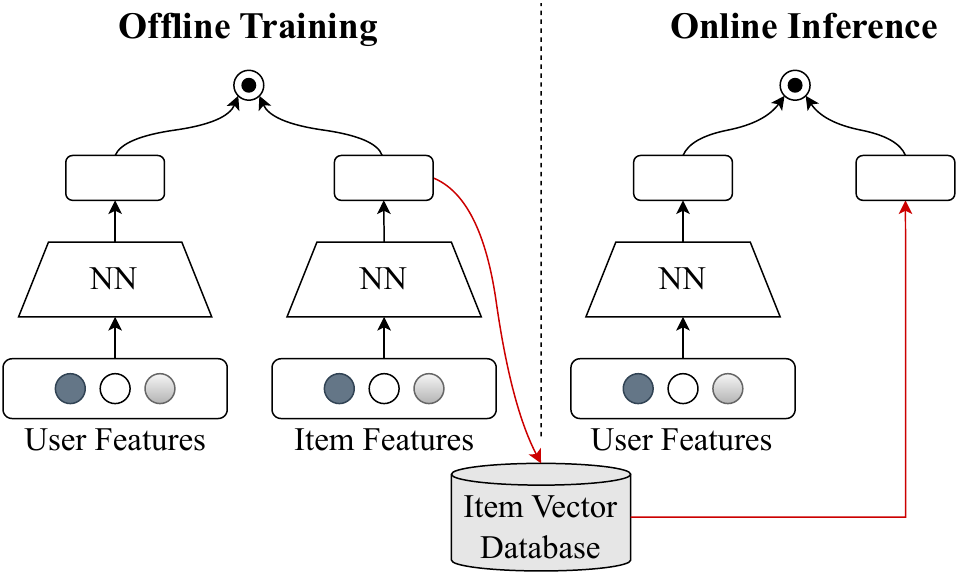}
        \label{fig:1b}
    }
    \caption{The architectures of Cascade Ranking System and Two-Tower Model}
    \label{fig:1}
\end{figure}

Large-scale search engines, advertisement systems, and recommendation systems are commonly employed in the modern Internet ecosystem, which facilitate information acquisition, commercial marketing, and personalized user experience. The core challenge is to quickly select a limited number of high-quality items from millions of candidates under strict latency constraints, in order to enhance user satisfaction and realize the platform's value. A multi-stage cascade ranking system\cite{chen2019behavior}\cite{fan2019mobius}\cite{qu2018product} is widely used to balance effectiveness and efficiency. As illustrated in Figure \ref{fig:1a}, the cascade ranking system is typically divided into four stages: matching, pre-ranking, ranking, and re-ranking. Each stage employs a model to score the candidate set output from the previous stage and selects a subset of candidates to pass on to the next stage, until the final few candidates are presented to the user. The initial stages usually perform a fast and rough ranking of a large number of candidates to ensure low inference latency, whereas the final stages use more complex and precise models to make accurate predictions.

Pre-ranking is a vital transitional stage that typically employs simple and efficient models to filter out a comparatively smaller set with higher potential relevance from a massive pool of candidates. The most commonly used pre-ranking model in the industry is the two-tower model\cite{huang2020embedding}\cite{huang2013learning}\cite{yang2020mixed}\cite{yu2021dual} (i.e, DSSM) as show in Figure \ref{fig:1b}, which designs a “user-item decoupling architecture” that utilizes neural networks to extract information within the user and item towers independently and then combines the user and item representations to make predictions in the output layer through operations like dot product, cosine similarity, etc. In order to reduce online computational consumption, the item representations can be pre-calculated offline and stored in the database for online retrieval service as the item features are mostly static or do not change frequently. During the online inference, benefiting from the user-item decoupling architecture, only the user representations need to be computed in real time, which provides excellent efficiency. However, limited by the parallel independent architecture, the learned latent representations of the two towers do not interact until the output layer, which is referred to as “\textit{Late Interaction}”\cite{khattab2020colbert}. Previous extensive research\cite{cheng2016wide}\cite{guo2017deepfm}\cite{wang2017deep}\cite{wang2021dcn} indicates that the interaction signals between user and item features have a significant impact on model performance, which is referred to as “\textit{Early Interaction}”. Though DAT\cite{yu2021dual} attempts to alleviate this issue by introducing augmented vectors to learn information from the opposite tower, the implicitly learned representations have limited expressiveness. COLD\cite{wang2020cold} and FSCD\cite{ma2021towards} achieve early interaction to improve the performance in a single-tower structure with several optimization tricks, but there are still concerns about inference efficiency. IntTower\cite{li2022inttower} and RankTower\cite{yan2024ranktower} introduce a sum-max similarity score in the output layer instead of dot product to enhance the interaction signals between the two towers, but it is unclear whether this hand-crafted reduction can capture arbitrary complex user-item interactions. Moreover, they lack sufficient early interactions and are not highly scalable due to additional computation and storage cost.

\begin{figure*}[t]
    \centering
    \includegraphics[width=\linewidth]{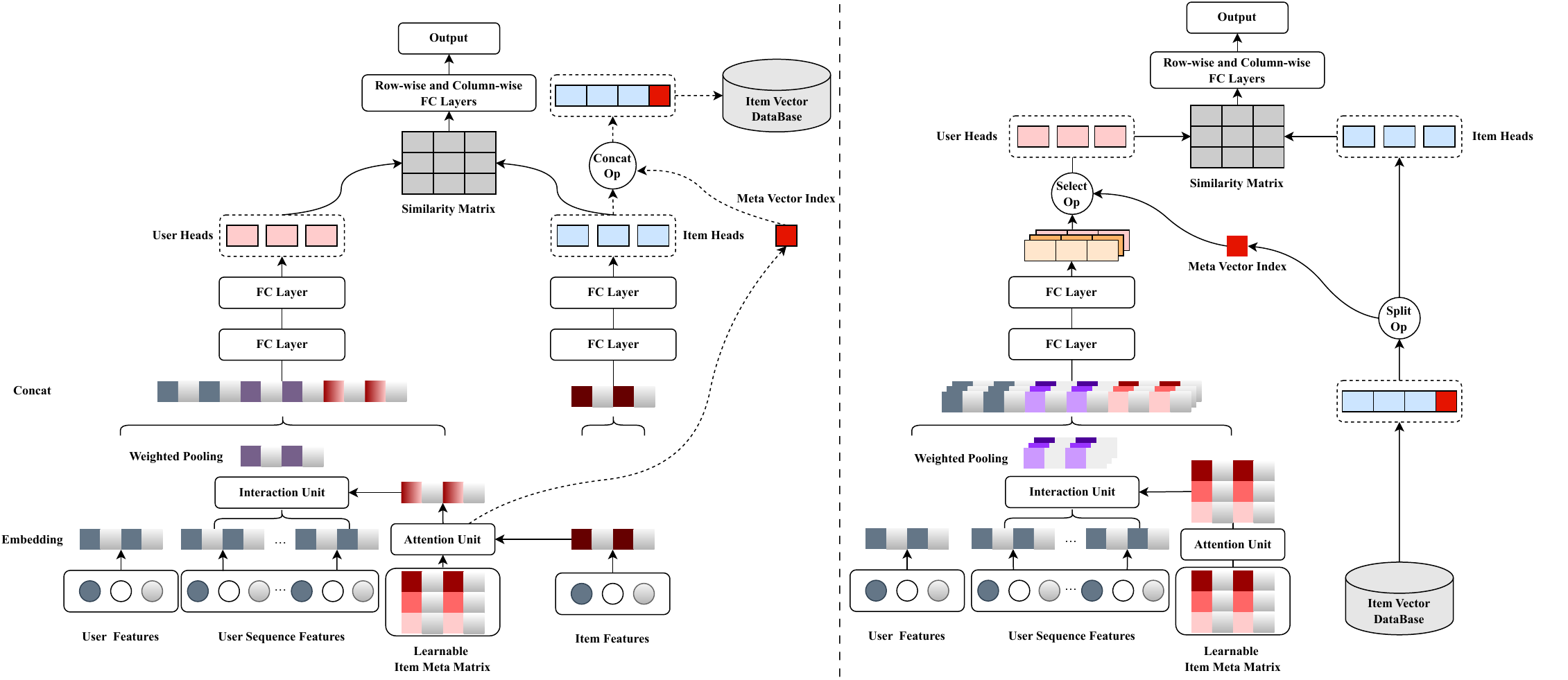}
    \caption{The overall architecture of our proposed FIT. The left part illustrates the training phase and the right part shows the inference phase.}
    \label{fig:2}
\end{figure*}

In this paper, we propose a novel architecture named \textit{learnable Fully Interacted Two-tower Model} (FIT), as shown in Figure \ref{fig:2}. We first design a Meta Query Module to customize a learnable item meta matrix to extract generalized information from the item tower. The user tower can utilize this matrix in the interaction unit to explicitly capture the expressive early user-item interaction signals and the model still follows the user-item decoupling architecture paradigm. Then a Lightweight Similarity Scorer is applied to further derive effective late interactions, which processes the similarity matrix between user and item representations via shallow fully connected layers. The scorer can be more accurate than the sum-max similarity score while having lower latency and storage cost.

Our main contributions are summarized as follows:
\begin{itemize}[leftmargin=1em]
    \item We design a Meta Query Module to make arbitrary interaction methods available while maintaining user-item decoupling architecture, which significantly enhances the early interactions between user and item towers explicitly.
    \item We introduce a Lightweight Similarity Scorer to effectively capture the late interaction signals to further improve the model performance. Moreover, LSS shows superiority in both effectiveness and efficiency over other late interaction structures.
    \item Extensive experiments on several public datasets demonstrate the effectiveness and efficiency of our proposed FIT over state-of-the-art methods. Further analyses also validate the effectiveness of the core designed techniques.
\end{itemize}

\section{RELATED WORK}
Pre-ranking is a strategic step aimed at achieving a balance between effectiveness and efficiency by quickly identifying the most promising candidates to be further ranked by the final ranking stages. LR\cite{mcmahan2013ad} and FM\cite{rendle2010factorization} are widely used in the age of shallow machine learning due to their simplicity and effectiveness. With the rapid development of deep learning\cite{covington2016deep}, the two-tower model (i.e., DSSM) based on neural networks has become the most popular pre-ranking model for large-scale recommendation systems, thanks to its strong representation learning ability and excellent efficiency. However, the user-item decoupling architecture, while bringing significant improvements in efficiency to meet the rapid response requirements of online recommendation systems, also notably reduces the interaction signals between user and item information, which in turn negatively impacts the model prediction accuracy. Therefore, an important research direction for pre-ranking models has been to enhance the interactions between user and item towers. Broadly, it can be divided into two aspects: early interaction methods and late interaction methods. The early interaction models improve the expressiveness of the model by enhancing the interactions between user and item features, while the late interaction models capture the interactive signals between user and item representations in the output layer.

DAT\cite{yu2021dual} is a typical early interaction two-tower model, which uses an augmented vector to capture the information from the other tower and regards the vector as the input feature of one tower. Although DAT allows for early interaction, the augmented vectors are not easy to learn and the expressiveness is limited. COLD\cite{wang2020cold} and FSCD\cite{ma2021towards} propose a single-tower model to fully achieve feature interactions with several optimization tricks for inference acceleration. However, the impact on efficiency is still difficult to ignore, especially in cases where the candidate set is large, as real-time model inference needs to be performed for each user-item pair. 

Inspired by ColBert\cite{khattab2020colbert}, IntTower\cite{li2022inttower} calculates a sum-max similarity score, instead of the commonly used dot product or cosine similarity score, between multi-layer user representations and last-layer item representation in the output layer. It contributes to improve the prediction accuracy while the user-item decoupling architecture enables high inference efficiency. Although IntTower utilizes the underlying information of the user tower when interacting with the item representation, it is still regarded as a late interaction model as it functions in the output layer and does not have explicit interactions between user and item features. As mentioned in LITE\cite{ji2024efficient}, it has been theoretically proved the restricted expressiveness of the sum-max score reduction. In addition, the storage cost of the multi-head item representations might also be a concern in recommendation systems with a large number of candidates. RankTower\cite{yan2024ranktower} extracts diversified latent representations of users and items, then a Gated Cross-Attention Network is applied to model bi-directional user-item interactions. The sum-max similarity score is calculated as the prediction output. Due to the introduction of sum-max similarity score, RankTower has similar issues to IntTower. Furthermore, the use of relatively complex attention structure in the output layer makes it difficult to balance online inference efficiency.

\section{PRELIMINARIES}
\label{sec:3}
Given 
\begin{math}
    \mathcal{U}=\{u_1,u_2,\ldots,u_m\}
\end{math}
and
\begin{math}
    \mathcal{V}=\{v_1,v_2,\ldots,v_n\}
\end{math}
as the set of users and items, we demote 
\begin{math}
    \mathcal{Y}\in\{0,1\}^{m\times n}
\end{math}
as the interaction matrix between users and items, where
\begin{math}
    y_{u,v}\in\mathcal{Y}
\end{math} 
indicates whether user 
\begin{math}
    u    
\end{math} 
gives an explicit positive feedback on item
\begin{math}
    v
\end{math}
, such as a purchase or click. 
\begin{math}
    \mathbf{e}_u
\end{math}
and
\begin{math}
    \mathbf{e}_v
\end{math}
are the concatenated feature vectors of user and item after the feature engineering procedure, in which usually continuous features are bucketized or scaled and categorical features are embedded into dense embedding vectors.
\begin{math}
    p(y=1|\mathbf{e}_u,\mathbf{e}_v)
\end{math}
is the predicted user-item interaction probability.

As illustrated in Figure \ref{fig:1b}, in general pre-ranking systems\cite{liu2017cascade}\cite{wu2018eenmf}, user features and item features are fed into the sub-towers separately, which are composed of FC (fully connected) layers with the ReLU\cite{glorot2011deep} activation function, to obtain corresponding representations in parallel. Formally, 
\begin{align}
    &\mathbf{h}^{i+1} = ReLU(\mathbf{W}^i\mathbf{h}^i+\mathbf{b}^i),\qquad i=0,1,\ldots,L-1 \\
    &\mathbf{h} = L2Norm(\mathbf{h}^L),
\end{align}
where 
\begin{math}
    \mathbf{W}^i \in \mathbb{R}^{d_{i+1}\times d_i}
\end{math}
and 
\begin{math}
    \mathbf{b}^i \in \mathbb{R}^{d_i}
\end{math}
denotes the weight matrix and bias vector for the 
\begin{math}
    i\mbox{-}th
\end{math}
layer, respectively. 
\begin{math}
    d_i
\end{math}
is the width of the 
\begin{math}
    i\mbox{-}th
\end{math}
FC layer and L is the depth of the FC layers.
\begin{math}
    \mathbf{h}^0 = \mathbf{e}
\end{math}
and 
\begin{math}
    d_0
\end{math}
is the dimension of the concatenated feature vectors. Representations 
\begin{math}
    \mathbf{h}_u
\end{math}
and
\begin{math}
    \mathbf{h}_v
\end{math}
are the output vectors of the L2 normalization layer after the 
\begin{math}
    L\mbox{-}th
\end{math}
FC layer in the user and item tower, respectively. Finally, the model output 
\begin{math}
    p_{uv}
\end{math}
is the dot product of the user and item representations:
\begin{equation}
    p_{uv}=\mathbf{h}_u^T\mathbf{h}_v.
    \label{eq:3}
\end{equation}
In general, user behaviors are relatively intensive, resulting in frequent dynamic changes in user features, while item features are mostly static or do not change frequently. To save time scoring a large number of candidates, the item representations
\begin{math}
    \mathbf{h}_v
\end{math}
are periodically pre-calculated and stored in the database, allowing for direct use in online real-time serving. During the pre-ranking stage in a user recommendation request, the user representations 
\begin{math}
    \mathbf{h}_u
\end{math}
are obtained by online model inference, and the 
\begin{math}
    k
\end{math}
candidate item representations
\begin{math}
    \mathbf{h}_i
\end{math}
can be retrieved from the database directly. Therefore, the overall time complexity of the pre-ranking two-tower model is 
\begin{math}
    O(N+kM),
\end{math}
where 
\begin{math}
    N
\end{math}
is the time taken for a neural network model inference and 
\begin{math}
    M
\end{math}
is the time consumed for item retrieval and scoring between each user-item representation pair. In contrast, the overall time complexity of the single-tower model is 
\begin{math}
    O(kN).
\end{math}
Since 
\begin{math}
    N
\end{math}
is much larger than
\begin{math}
    M,
\end{math}
the two-tower model shows efficiency superiority over the single-tower model.

It should be noted that the user and item representations do not interact until the output layer, which is referred to “Late Interaction”. Some commonly used and effective interaction methods cannot be applied in this structure, causing the model effect to be significantly weakened. For example, user behavior sequences are often very important features and many proven network can be used to extract effective information from the interaction of target item and user sequence features. However, without the opposite tower information, user sequence features can only be roughly processed in the two-tower model using methods like sum/average pooling. Moreover, the limited expressiveness of the final dot product leads to insufficient late interaction between the two towers. We propose FIT to mitigate the above issues with enhanced early and late interactions.

\section{THE PROPOSED MODEL}
In this section, we will provide a detailed description of our proposed FIT model, which achieves both expressive early interaction between user and item features and effective late interaction between user and item towers. The overall architecture is illustrated in Figure \ref{fig:2}, which still follows the two-tower decoupling paradigm. FIT mainly consists of two components: MQM (Meta Query Module) and LSS (Lightweight Similarity Scorer). MQM introduces a learnable item meta matrix to extract generalized information from the item tower. The user tower then utilizes this matrix to explicitly capture the user-item interaction signals in the interaction unit, where arbitrary interaction networks can be applied as in the single-tower structure, such as DeepFM\cite{guo2017deepfm}, DIN\cite{zhou2018deep}, DIEN\cite{zhou2019deep}, CAN\cite{bian2022can}, etc. The typical DIN is introduced here to make full use of the user behavior sequence features. LSS is designed to replace the commonly used dot product in the output layer to strengthen the late information interaction between the two towers, which processes the similarity matrix between user and item representations via shallow FC layers. Compared to the sum-max similarity score, which is frequently introduced in recent pre-ranking studies\cite{li2022inttower}\cite{yan2024ranktower}, LSS is theoretically demonstrated to be more expressive.

\begin{figure}[t]
  \centering
  \includegraphics[width=\linewidth]{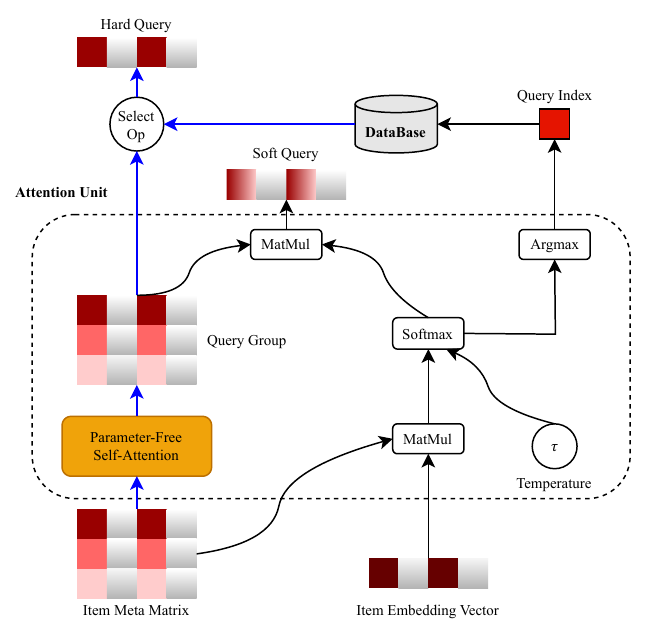}
  \caption{The architecture of Meta Query Module, where the blue arrows denote the inference flow.}
  \label{fig:3}
\end{figure}

\subsection{Meta Query Module}
\label{sec:4.1}
As shown in Figure \ref{fig:3}, we define an embedding matrix 
\begin{math}
    \mathbf{Q} \in \mathbb{R}^{N \times D}
\end{math}
named “\textit{item meta matrix}”, where 
\begin{math}
    N
\end{math}
is a hyper-parameter that denotes the item meta vector size and 
\begin{math}
    D
\end{math}
is the embedding dimension equal to the input of the attention unit. The item meta matrix is trainable and represents generalized information from the item tower. It is recommended to initialize the learnable parameters of the matrix using a uniform distribution, which provides better numerical stability and performs better in practice. The input of the attention unit 
\begin{math}
    \mathbf{e_c} \in \mathbb{R}^{D}
\end{math}
is the concatenated embedding vector of the item attribute feature set
\begin{math}
    \{\mathbf{e}_{i1},\mathbf{e}_{i2},\ldots\}
\end{math}
, such as item ID, category ID, etc.
\begin{equation}
    \mathbf{e}_c = [\mathbf{e}_{i1}||\mathbf{e}_{i2}||\ldots],
\end{equation}
where 
\begin{math}
    ||
\end{math}
denotes the vector concatenation operation,
\begin{math}
    D=d\times l,
\end{math}
\begin{math}
    d
\end{math}
is the item feature embedding dimension and 
\begin{math}
    l
\end{math}
is the size of the item attribute feature set.

Parameter-free self-attention, a simplified implementation of the scaled dot-product attention similar to some related works\cite{yang2021simam}\cite{zhai2023simple}, is adopted to accelerate parameter convergence of the item meta matrix. Instead of learning the projection matrices (typically denoted as 
\begin{math}
    \mathbf{W}_Q,\mathbf{W}_K,
\end{math}
and
\begin{math}
    \mathbf{W}_V)
\end{math}
as traditional attention mechanism, the queries, keys, and values are directly derived from the input, i.e. the item meta matrix. Formally, the output item meta vector group
\begin{math}
    \mathbf{Q}^* \in \mathbb{R}^{N \times D}
\end{math}
, referred to as the “\textit{query group}”, can be calculated as follows:
\begin{equation}
    \mathbf{Q}^*=(\mathbf{Q}\mathbf{Q}^T)\mathbf{Q}.
\end{equation}
The item meta vector
\begin{math}
    \mathbf{q}_i^* \in \mathbb{R}^D,
\end{math}
referred to as the “\textit{hard query}”, 
is the 
\begin{math}
    \textit{i}\mbox{-}th
\end{math}
row of the query group
\begin{math}
    \mathbf{Q}^*
\end{math}
and is considered to be the aggregated representation of the clustered item set. It is able to achieve explicit early interaction between user and item by using the hard query in the user tower. It should be noted that the query group can be calculated directly without item attribute features in the inference phase. Therefore, FIT still follows the two-tower decoupling paradigm, maintaining efficient online performance.

Next, the softmax method is used to obtain similarity scores between the candidate item and the query group. The index of the hard query with the maximum similarity score, named query index, is stored together with the offline pre-calculated item representation and will be used to select the corresponding hard query of the candidate item in the online prediction service. The index only occupies 1 bit of the stored vector and has a negligible impact on storage costs. If only the associated hard query of the candidate item is involved in training, then the parameters of the hard query for each user-item pair will be the only ones updated. This is not conducive to learning the parameters of the item meta matrix, especially when the items are sparse. In order to speed up the convergence of the item meta matrix, we use the weighted item meta vector 
\begin{math}
    \mathbf{\tilde{q}^*} \in \mathbb{R}^{D}
\end{math}
, referred to as the “\textit{soft query}”, to achieve early interaction in the training phase. In addition, for the sake of consistency between training and inference, we use the temperature mechanism in the softmax method for smooth control. The weight vector, also known as the similarity scores,
\begin{math}
    \mathbf{k} \in \mathbb{R}^{N}
\end{math}
is calculated as follows:
\begin{equation}
    k_i = \frac{exp(\mathbf{e}_c \mathbf{q}_i^T/\tau)}{\sum_{t=1}^{N}exp(\mathbf{e}_c \mathbf{q}_t^T/\tau)},
\end{equation}
the soft query
\begin{math}
    \mathbf{\tilde{q}^*}  
\end{math}
and the query index
\begin{math}
    s
\end{math}
can be represented as:
\begin{equation}
    \mathbf{\tilde{q}^*} = \mathbf{k}\mathbf{Q}^*,
\end{equation}
\begin{equation}
    s = \arg\max_{1\leq i \leq N} k_{i},
\end{equation}
where 
\begin{math}
    \mathbf{q}_i \in \mathbb{R}^D
\end{math}
is the 
\begin{math}
    i\mbox{-}th
\end{math}
row of the meta matrix
\begin{math}
    \mathbf{Q}
\end{math}
and 
\begin{math}
    \tau
\end{math}
is the temperature parameter, which linearly decays from 1.0 to a number slightly larger than 0 as the global training step increases and then remains constant. When the temperature parameter is around 1.0, the similarity scores after softmax are relatively uniform, allowing multiple hard queries to be simultaneously updated, which accelerates the convergence of the item meta matrix. As the temperature parameter gradually approaches 0, the hard query corresponding to the maximum similarity score mainly takes effect, and almost only this query vector is updated, thus maintaining consistency with the inference phase. The soft query and hard query are utilized in training and inference phase, respectively. We give a detailed description in Section \ref{sec:4.3}.

In summary, MQM enhances the model's expressiveness by generating learnable clustered item representations that support explicit early user-item interactions. Moreover, it accelerates the convergence of the item meta matrix while maintaining consistency between training and inference using parameter-free self-attention and a softmax weighting technique with a decaying temperature parameter. Furthermore, the model retains the two-tower decoupling paradigm and introduces almost no additional parameters, ensuring efficient performance.

\subsection{Lightweight Similarity Scorer}
\label{sec:4.2}
The dot product as in Eq.\ref{eq:3} is well-regarded in the two-tower output layer due to its simplicity and efficiency, yet it also hinders the model's expressiveness because of late interactions. To alleviate this problem, one may seek to apply MLP to calculate the output score as follows:
\begin{equation}
    p_{uv} = MLP([\mathbf{h}_u||\mathbf{h}_v]).
\end{equation}
However, as demonstrated in \cite{rendle2020neural}, while MLP can in theory approximate any function, it is non-trivial to learn a dot product with a single-layer MLP, while muti-layer MLP is too costly to use for online recommendation. In order to enhance the late interactions between the two towers, the sum-max similarity score is introduced in the output layer in recent pre-ranking studies\cite{li2022inttower}\cite{yan2024ranktower}. Formally, the score is calculated as follows:
\begin{equation}
    p_{uv} = \sum\limits_{i=1}^{H_u}\mathop{\max}_{j\in\{1,\ldots,H_v\}}(\mathbf{z}_u^i)^T\mathbf{z}_v^j,
\end{equation}
where 
\begin{math}
    \mathbf{z}_u^i
\end{math}
is the 
\begin{math}
    i\mbox{-}th
\end{math}
sub-space of the user representation, similarly 
\begin{math}
    \mathbf{z}_v^j
\end{math}
is the 
\begin{math}
    j\mbox{-}th
\end{math}
sub-space of the item representation. Though the sum-max similarity score achieves better accuracy than dot product, it is unclear whether this hand-crafted reduction can capture arbitrary complex user-item interactions. Moreover, even though the individual vectors involved in the calculation become smaller, it might have higher latency than dot product: calculating the similarity matrix 
\begin{math}
    \mathbf{z}_u^T\mathbf{z}_v
\end{math}
requires 
\begin{math}
    H_u \cdot H_v
\end{math}
dot products. In addition, as mentioned in Section \ref{sec:3}, the item representations are pre-calculated offline and stored in the database for online retrieval. The storage cost of the multi-head item representations might also be a concern in recommendation systems with a large number of candidates. Inspired by LITE\cite{ji2024efficient}, we propose a Lightweight Similarity Scorer to achieve effective late interactions between the two towers with superiority in both effectiveness and efficiency.

We first map user/item representations into muti-head sub-spaces to extract diverse information. The user and item representations can be mapped into 
\begin{math}
    H_u
\end{math}
and
\begin{math}
    H_v
\end{math}
sub-spaces, respectively. The 
\begin{math}
    h\mbox{-}th
\end{math}
sub-space of the user representation 
\begin{math}
    \mathbf{z}_u^h \in \mathbb{R}^p
\end{math}
is obtained from:
\begin{equation}
    \mathbf{z}_u^h = \mathbf{W}_u^h\mathbf{h}_u^L+\mathbf{b}_u^h, \qquad h=1,2,\ldots,H_u
\end{equation}
where 
\begin{math}
    \mathbf{W}_u^h \in \mathbb{R}^{d_L \times p}
\end{math}
and
\begin{math}
    \mathbf{b}_u^h \in \mathbb{R}^{p}
\end{math}
denotes the weight matrix and bias vector for the 
\begin{math}
    h\mbox{-}
\end{math}
head sub-space respectively, 
\begin{math}
    p
\end{math}
is the dimension of sub-space. It is worth noting that the
\begin{math}
    H_u
\end{math}
head transformation can be calculated in parallel and the stacked result is 
\begin{math}
    \mathbf{z}_u \in \mathbb{R}^{p \times H_u}.
\end{math}

Similarly, the
\begin{math}
    h\mbox{-}th
\end{math}
sub-space of the item representation
can be represented as:
\begin{equation}
    \mathbf{z}_v^h = \mathbf{W}_v^h\mathbf{h}_v^L+\mathbf{b}_v^h, \qquad h=1,2,\ldots,H_v
\end{equation}
and the stacked result is 
\begin{math}
    \mathbf{z}_v \in \mathbb{R}^{p \times H_v}.
\end{math}

Then, the similarity matrix
\begin{math}
    \mathbf{S} \in \mathbb{R}^{H_u \times H_v},
\end{math}
which consists of the dot products of all user-item head pairs, can be calculated as follows:
\begin{equation}
    S_{ij} = (\mathbf{z}_u^i)^T\mathbf{z}_v^j.
\end{equation}
Next, the row-wise and column-wise FC layers are conducted on the similarity matrix. Specifically, we first apply row-wise updates to
\begin{math}
    \mathbf{S},
\end{math}
then column-wise updates, and finally a linear projection to get a scalar score. Formally, we calculate
\begin{math}
    \mathbf{S}^{\prime} \in \mathbb {R}^{H_u \times d}
\end{math}
and
\begin{math}
    \mathbf{S}^{\prime\prime} \in \mathbb{R}^{d\times d}.
\end{math}
For all 
\begin{math}
    1\leq i \leq H_u
\end{math}
and
\begin{math}
    1\leq j \leq H_v
\end{math}
,
\begin{equation}
    \mathbf{S}_{i,:}^{\prime} = ReLU(\mathbf{W}_1\mathbf{S}_{i,:}+\mathbf{b}_1),
\end{equation}
\begin{equation}
    \mathbf{S}_{:,j}^{\prime\prime} = ReLU(\mathbf{W}_2\mathbf{S}_{:,j}^{\prime}+\mathbf{b}_2),
\end{equation}
where
\begin{math}
    d
\end{math}
is the feature embedding dimension,
\begin{math}
    \mathbf{W}_1 \in \mathbb{R}^{d\times H_u}
\end{math}
and
\begin{math}
    \mathbf{b}_1 \in \mathbb{R}^{H_u}
\end{math}
are the row-wise weight matrix and bias vector,
\begin{math}
    \mathbf{W}_2 \in \mathbb{R}^{d\times d}
\end{math}
and
\begin{math}
    \mathbf{b}_2 \in \mathbb{R}^{d} 
\end{math}
are the column-wise weight matrix and bias vector. The output score is calculated as:
\begin{equation}
    p_{uv} = \mathbf{w}^T Flatten(\mathbf{S}^{\prime\prime}),
\end{equation}
where 
\begin{math}
    \mathbf{w} \in \mathbb{R}^{d^2}
\end{math}
is the projection vector and 
\begin{math}
    Flatten
\end{math}
reshapes the matrix
\begin{math}
    \mathbf{S}^{\prime\prime}
\end{math}
into a flattened vector.

Notably, there is another way to calculate the output score: flatten
\begin{math}
    \mathbf{S}
\end{math}
and then apply FC layers as follows:
\begin{equation}
    p_{uv} = (\mathbf{w}^{\prime})^T(\mathbf{W}_3Flatten(\mathbf{S}) + \mathbf{b}_3),
\end{equation}
where
\begin{math}
    \mathbf{W}_3 \in \mathbb{R}^{d\times H_u H_v}
\end{math}
and
\begin{math}
    \mathbf{b}_3 \in \mathbb{R}^{H_u H_v}
\end{math}
are the weight matrix and vias vector, and
\begin{math}
    \mathbf{w}^{\prime} \in 
\end{math}
is the projection vector. Obviously, the row-wise and column-wise FC layers have fewer parameters, which can reduce the computational cost of online inference, especially when the head size 
\begin{math}
    H_u
\end{math}
and 
\begin{math}
    H_v
\end{math}
are large. 

Furthermore, it has been demonstrated in detail as in \cite{ji2024efficient} that the row-wise and column-wise similarity scorer is a universal approximator of continuous scoring functions in 
\begin{math}
    \ell_2
\end{math}
distance, even under tight head size constraints. Therefore, LSS can approach the theoretical bound using only a small number of parameters, enhancing late interactions while maintaining online inference efficiency.

\subsection{Model Training and Inference}
\label{sec:4.3}
In this section, we will show in detail how our proposed FIT works during training and inference. 
\subsubsection{Training}
As mentioned in Section \ref{sec:4.1}, the soft query that represents generalized information from the item tower is produced to achieve early interaction with user information in the training phase. Arbitrary interaction methods can be applied in the interaction unit to combine the user and item information. Concretely, a typical DIN\cite{zhou2018deep} is utilized to capture user's diverse interests effectively from historical behaviors, which is difficult to perform effectively in the traditional two-tower model due to the lack of item information. The embedded representation of the user historical behavior sequence features is defined as a list of embedding vectors:
\begin{math}
    \{\mathbf{e}_u^{i_1},\mathbf{e}_u^{i_2},\ldots,\mathbf{e}_u^{i_K}\},
\end{math}
where 
\begin{math}
    K
\end{math}
is the sequence length. It is a common practice\cite{cheng2016wide}\cite{covington2016deep} to transform the list of embedding vectors via a sum or average pooling layer to get a fixed-length vector:
\begin{equation}
    \mathbf{e}_u^i = pooling(\mathbf{e}_u^{i_1},\mathbf{e}_u^{i_2},\ldots,\mathbf{e}_u^{i_K}).
\end{equation}
In contrast, a weighted pooling vector can be obtained in FIT.
\begin{equation}
    \mathbf{e}_u^i = f(\mathbf{\tilde{q}^*}, \mathbf{e}_u^{i_1},\mathbf{e}_u^{i_2},\ldots,\mathbf{e}_u^{i_K})=\sum\limits_{j=1}^{K}a(\mathbf{e}_u^{i_j},\mathbf{\tilde{q}^*})\mathbf{e}_u^{i_j},
\end{equation}
where 
\begin{math}
    a(\cdot)
\end{math}
is a feed-forward network with output as the activation weight. The weighted pooling vector and the soft query are concatenated with other user feature vectors. 
\begin{equation}
    \mathbf{e}_u = [\mathbf{e}_u^1||\ldots||\mathbf{e}_u^i||\ldots||\mathbf{e}_u^A||\mathbf{\tilde{q}^*}],
\end{equation}
where 
\begin{math}
    A
\end{math}
is the user features number. The concatenated user feature vectors 
\begin{math}
    \mathbf{e}_u
\end{math}
is then fed into the multi FC layers to obtain the user representations. The final output score is calculated as in Section \ref{sec:4.2}. The cross-entropy loss is employed as follows:
\begin{equation}
    loss = -\frac{1}{T}\sum\limits_{(u,v,y)\in\mathcal{T}}(ylog\sigma(p_{uv})+(1-y)log(1-\sigma(p_{uv}))),
\end{equation}
where 
\begin{math}
    T
\end{math}
is the number of user-item pairs in the training dataset 
\begin{math}
    \mathcal{T}
\end{math}
and 
\begin{math}
    \sigma(\cdot)
\end{math}
denotes the sigmoid function.

\subsubsection{Inference}
Different from the training phase, the hard query is utilized to interact with user information in the inference phase. The query index 
\begin{math}
    s
\end{math}
, which is stored in the database together with the item representation
\begin{math}
    \mathbf{z}_{v},
\end{math}
is split from the stored concatenated vector
\begin{math}
    [s||\mathbf{z}_{v}]
\end{math}
to find the hard query corresponding to the candidate item. In order to highly decouple the calculation of user and item representations, all the hard queries in the query group
\begin{math}
    \mathbf{Q}^*
\end{math}
participate in the user tower inference, which can be computed in parallel. The stacked result is
\begin{math}
    \{\mathbf{z}_u^{(1)},\mathbf{z}_u^{(2)},\ldots,\mathbf{z}_u^{(N)}\} \in \mathbb{R}^{p\times H_u\times N}.
\end{math}
When calculating the output score, only one user representation identified by the query index will be selected to obtain the similarity matrix ensuing online inference efficiency.

\section{EXPERIMENTS}
In this section, we conduct thorough experiments to evaluate our proposed FIT on three publicly available real-word datasets. Experiments are described in detail, including datasets, evaluation metrics, comparisons with state-of-the-art pre-ranking models, and corresponding analyses. We aim to answer the following research questions.

\begin{itemize}[leftmargin=1em]
    \item
    \begin{math}
        \mathbf{RQ1.}
    \end{math}
    How does our proposed model perform compared with state-of-the-art pre-ranking models in terms of effectiveness and efficiency?
    \item 
    \begin{math}
        \mathbf{RQ2.}
    \end{math}
    How much does each of our proposed components contribute to the model?
    \item 
    \begin{math}
        \mathbf{RQ3.}
    \end{math}
    How do key hyper-parameters influence the performance?
\end{itemize}

\subsection{Experimental Settings}
\subsubsection{Datasets}
We evaluate our proposed model employing three popular public available real-world datasets and construct user behavior sequence features on all the datasets to better evaluate the model's interaction capability.
\begin{itemize}[leftmargin=1em]
    \item 
    \begin{math}
        \mathbf{Amazon\,Electronics}\footnote{\url{http://jmcauley.ucsd.edu/data/amazon/}\label{sharedfootnote}}\mathbf{(Amazon\,Electro)}
    \end{math}
    The subset of the Amazon dataset named Electronics is utilized in our experiments. It contains 192,403 users, 63,001 goods, and 1,689,188 reviews. As the previous works\cite{zhou2019deep}\cite{zhou2018deep} did, we regard reviews as behaviors and sort the reviews from one user by time. The task is to predict whether the user will write reviews that shown in 
    \begin{math}
        T\mbox{-}th
    \end{math}
    review by making use of the preceding 
    \begin{math}
        T-1
    \end{math}
    behaviors as the user behavior sequence feature, given all
    \begin{math}
        T
    \end{math}
    behaviors truncated at a maximum length of 50. We take the
    \begin{math}
        T\mbox{-}th
    \end{math}
    reviewed good as positive sample and randomly sample an unreviewed good as negative sample.

    \item 
    \begin{math}
        \mathbf{Amazon \,Books}\textsuperscript{\ref{sharedfootnote}}
    \end{math}
    We use another subset of the Amazon dataset named Books, which contains 603,668 users, 367,982 goods and 8,898,041 reviews. The same processing method as Amazon Electro is applied.
    
    \item 
    \begin{math}
    \mathbf{MovieLens\mbox{-}1M}\footnote{\url{https://grouplens.org/datasets/movielens/}}\mathbf{(ML\mbox{-}1M)}
    \end{math}
    This dataset consists of 1,000,000 movie ratings provided by 6,040 users and 3,668 movies. We regard ratings as behaviors and handle them similarly to how we process Amazon Electro.

    \item 
    \begin{math}
        \mathbf{Taobao\,Display\,Ad\,Click}\footnote{\url{https://tianchi.aliyun.com/dataset/56}}\mathbf{(Taobao)}
    \end{math}
    It comprises 26,557,961 interactions between 1,061,768 users and 785,597 advertisements on Taobao website. We collect users' historical click behaviors prior to the exposure time of each pageview record to construct the behavior sequence feature. Each sequence is truncated to a maximum length of 50, and users without any click behaviors are excluded.
\end{itemize}

\subsubsection{Evaluation Metrics.} We adopt three widely used metrics to evaluate the performance of the models.
\begin{itemize}[leftmargin=1em]
    \item 
    \begin{math}
        \mathbf{AUC}
    \end{math}
    It evaluates the performance of a binary classification model by quantifying the area under the ROC curve. An AUC value closer to 1 indicates better performance.

    \item 
    \begin{math}
        \mathbf{Logloss}
    \end{math}
    This metric measures the probability assigned by the model to the actual observed class and reflects the uncertainty between the predicted probabilities and the actual labels. A lower value suggests more accurate model predictions.

    \item 
    \begin{math}
        \mathbf{RelaImpr} 
    \end{math}
    It is introduced to measure the relative improvement over the models and is defined as below:
    
    \begin{equation}
        RelaImpr = (\frac{AUC(measured\,model)-0.5}{AUC(base\,model)-0.5}-1)\times100\%.
    \end{equation}
\end{itemize}

\subsubsection{Competing Models.} We choose the following pre-ranking models for comparisons: \textbf{Two-Tower}\cite{huang2013learning}, \textbf{DAT}\cite{yu2021dual}, \textbf{COLD}\cite{wang2020cold}, \textbf{IntTower}\cite{li2022inttower}, \textbf{RankTower}\cite{yan2024ranktower}.

\subsubsection{Implementation Details.} All models are implemented with PyTorch\cite{paszke2019pytorch}. The mini-batch size is 2048, and the embedding dimension is 16 for all the categorical features. We employ Adam\cite{kingma2014adam} optimizer with learning rate set to 0.001. Batch Normalization\cite{ioffe2015batch} is applied to avoid overfitting. The depth of the FC layers
\begin{math}
    L
\end{math}
is set to 3 and the number of hidden units is set as [300, 300, 128] with ReLU as the activation function. The other hyper-parameters of each model are grid searched to obtain the best results individually, ensuring the model comparison fair. We save all models that perform best on the validation set which is 20 percent randomly selected from the training set during training, and load the saved models to predict results on the test set. We conduct experiments with Tesla A100 GPUs.

\begin{table*}[t]
  \caption{Performance comparison of different models on ML-1M, Amazon Electro, Amazon Books and Taobao (The bold font indicates the optimal metric, and the underline represents the suboptimal metric). “*” indicates the statistically significant improvements (i.e., two-sided t-test with $p < 0.05$) over the best baseline.}
  \label{tab:1}
    \resizebox{\textwidth}{!}{%
  \begin{tabular} {ccccccccccccc}
    \hline
     \multirow{2}{*}{Model} & \multicolumn{3}{c}{ML-1M} & \multicolumn{3}{c}{Amazon Electro} & \multicolumn{3}{c}{Amazon Books} & \multicolumn{3}{c}{Taobao}\\
     \cmidrule(r) {2 -13}
     &  AUC & Logloss & RelaImpr  & AUC & Logloss & RelaImpr& AUC & Logloss & RelaImpr& AUC & Logloss & RelaImpr \\  \hline
     Two-Tower & 0.8695  & 0.4508  & 0.00\%  & 0.8441 & 0.4870  & 0.00\% & 0.8742&	0.4484& 0.00\% & 0.6577 & 0.2257 & 0.00\% \\   
     DAT & 0.8784 & 0.4363  & 2.41\%	& 0.8465 & 0.4858 & 0.70\% &0.8802&	0.4413&	1.61\%&0.6605 & 0.2241 & 1.74\%\\ 
     COLD & 0.9009 &	0.4316 & 8.51\%	&0.8504 &  0.4833 & 1.82\% &0.8996&\underline{0.4072}&6.80\%&\underline{0.6645} & \underline{0.2236} & 4.28\%\\
     IntTower & 0.9036 & 0.4142 & 9.24\%	& 0.8490 & \underline{0.4807} & 1.42\% & 0.8972&0.4444&6.15\%	 & 0.6638	&0.2245	&3.81\%\\
     RankTower &\underline{0.9095} &	\underline{0.3986} &10.82\%	& \underline{0.8520} & 0.4797  &2.31\%& \underline{0.9031}&	0.4149	&7.75\%& 0.6630 &	0.2250 & 3.34\%\\  \hline
     FIT & \textbf{\,\,0.9225*} & \textbf{\,\,0.3766*} & \textbf{14.34\%}& \textbf{\,\,0.8598*} & \textbf{\,\,0.4707*} & \textbf{4.58\%} & \textbf{\,\,0.9171*}& \textbf{\,\,0.3731*} &\textbf{11.47\%} & \textbf{\,\,0.6736*}& 	\textbf{\,\,0.2226*}  &	\textbf{10.03\%}\\        \bottomrule
     \end{tabular}
     }
\end{table*} 

For FIT, we use Kaiming uniform initialization\cite{he2015delving} to initialize the item meta matrix and the decaying temperature parameter
\begin{math}
    \mathbf{\tau}
\end{math}
is set to:
\begin{equation}
    \tau = max(min(1.0, 1-\text{training\_steps}/\text{threshold}), 0.001),
\end{equation}
where \text{threshold} is the number of training steps in an epoch for each dataset. The size of the item meta vectors 
\begin{math}
    N
\end{math}
is set to 64. The number of heads
\begin{math}
    H_u,H_v
\end{math}
in LSS is set to 2 with dimension 64. DIN is applied in the interaction unit to make full use of the user behavior sequence features. The number of the attention hidden units is set as [64, 16, 1] and the attention activation function is Dice\cite{zhou2018deep}.

\subsection{Comparison of Model Effectiveness (RQ1)}
\label{sec:5.2}
Table \ref{tab:1} shows the performance of FIT and the baseline models on ML-1M, Amazon Electro, Amazon Books, and Taobao. All experiments are repeated 10 times to fairly obtain the averaged results with comparable computational costs.

Obviously, all models with specially designed structures beat Two-Tower significantly, which demonstrates the importance of user-item interactions. Specifically, DAT and COLD gain a performance boost through early interaction, while IntTower and RankTower achieve improvement by late interaction, indicating that both early interaction and late interaction contribute to enhanced performance. DAT designs an implicit early interaction structure between the two towers with limited interaction capability. COLD applies the single-tower structure and achieves strong performance with early interaction between user and item, which requires extra feature engineering processes and optimization tricks for online inference acceleration while the impact on inference efficiency cannot be overlooked. IntTower achieves performance improvement by replacing dot product with the sum-max similarity score in the output layer. RankTower implements a late interaction with richer expressiveness via the cross-attention network.

FIT significantly outperforms all baseline models on four datasets. It demonstrates that FIT improves the restricted performance of the two-tower architecture by expressive early interaction and effective late interaction. Notably, the performance of FIT does not stop here. To ensure a fairer comparison of results, FIT choose hyper-parameters combination with relatively fewer parameters. Nevertheless, the improvements achieved are still quite evident. More detailed analyses of the impact of hyper-parameters will be provided in Section \ref{sec:5.5}.

\begin{figure}[t]
    \centering
    %%\setlength{\abovecaptionskip}{0.2cm}
    %%\setlength{\belowcaptionskip}{-0.5cm}
    % \subfigure[Amazon Books] {
        \includegraphics[width=1\linewidth]{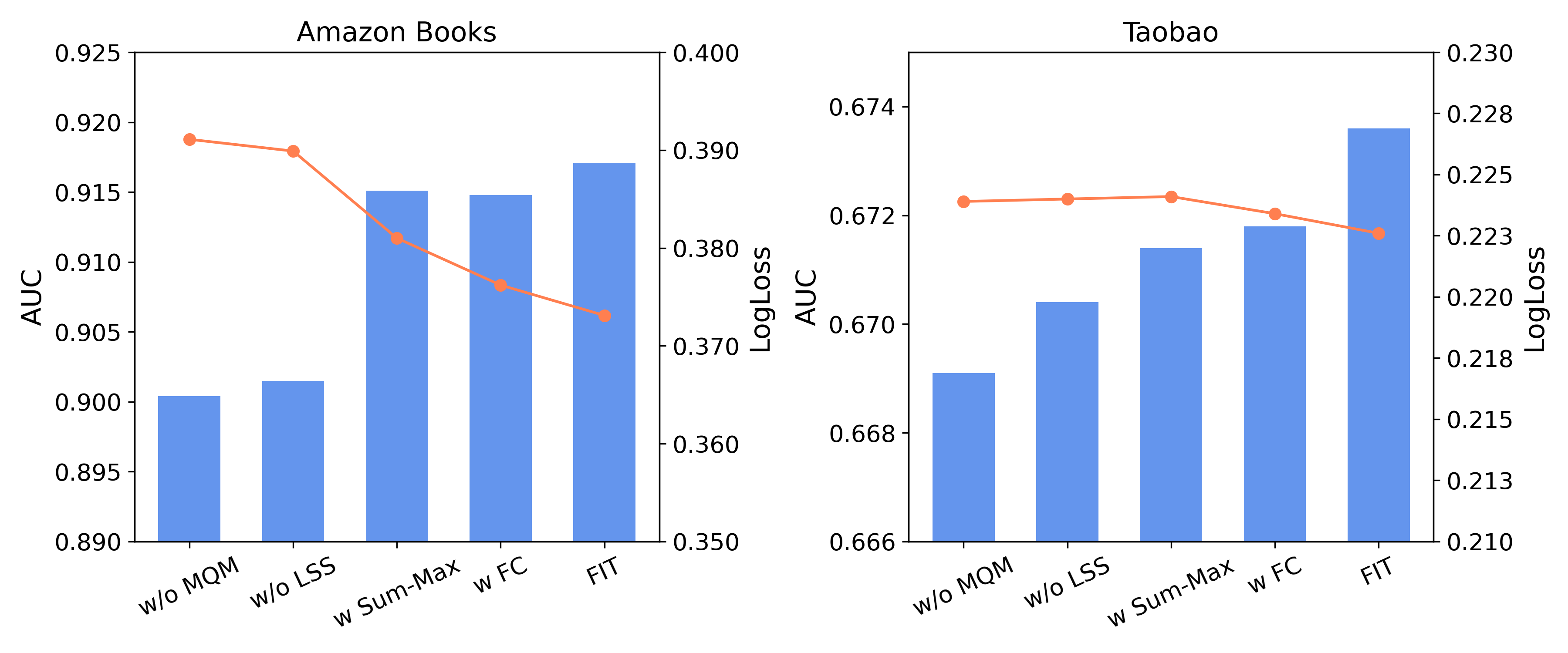}
        \label{fig:4a}
    % }
    % \subfigure[Taobao] {
    %     \includegraphics[width=0.45\linewidth]{ablation.png}
    %     \label{fig:4b}
    % }
    \caption{The component effectiveness analysis of FIT on the two datasets.}
    \label{fig:4}
\end{figure}

\subsection{Comparison of Model Efficiency (RQ1)}
In pre-ranking systems, it is crucial to focus on model efficiency, which typically involves optimizing training and inference efficiency, as well as minimizing the storage cost of the item representations. Specifically, training efficiency affects the model update frequency, which is vital for maintaining up-to-date recommendations. Frequent updates also help mitigate the cold start problem for new items. During real-time requests, thousands of items need to be scored in the pre-ranking stage, which is dozens of times greater than in the ranking stage. Models are required to make rapid predictions. In addition, item representations are typically pre-calculated offline to reduce the real-time computational burden. These pre-calculated representations are stored in a database for fast retrieval during online inference. Efficient storage reduces costs and decreases memory usage on the serving machines, improving overall system performance.
\begin{table}[t]
  \caption{Efficiency comparison of different models on Taobao in terms of training time in an epoch, pre-ranking inference latency for one request with 2000 candidates, and storage cost of item representations.}
  \label{tab:2}
  \begin{tabular} {cccc}
    \hline
    \textbf{Model} & \textbf{Training Time} & \textbf{Latency} & \textbf{Storage} \\
    \hline
    Two-Tower & 228\,s & 25.4\,ms & 1\,$\times$ \\
    DAT & 238\,s & 25.6\,ms & 0.25\,$\times$ \\
    COLD & 284\,s & 31.4\,ms & 0\,$\times$ \\
    IntTower & 280\,s & 27.4\,ms & 8\,$\times$ \\
    RankTower & 231\,s & 33.3\,ms & 1\,$\times$ \\
    \hline
    FIT & 260\,s & 26.9\,ms & 1\,$\times$ \\
    \hline
     \end{tabular}
\end{table} 

Table \ref{tab:2} shows the training time, inference latency, and storage cost of different models on the largest dataset Taobao. All experiments are repeated 10 times and averaged results are reported. As described in Section \ref{sec:3}, for models except COLD with two-tower structure, the item representations are pre-stored in the database and the inference latency consists of two parts: user representation inference and user-item pair scoring. In contrast, for COLD with single-tower structure, the inference latency involves both user and item representations inference as well as user-item pair scoring. Due to the introduction of two corresponding augmented vectors by user and item IDs in DAT, which have the same dimensions as the final representation vectors, the number of parameters significantly increase, making the model prone to overfitting and greatly reducing training efficiency. Therefore, to improve the model's performance, the item representation size and storage cost of DAT is smaller compared to other two-tower structure models. COLD with single-tower structure has no storage cost, but the computation cost is high. IntTower introduces multi-head representations, which bring additional computation and storage cost. RankTower employs a relatively complex multi-head cross-attention network in the output layer, resulting in the highest inference latency. Although FIT utilizes row-wise and column-wise FC layers with a small number of parameters in the output layer, it can achieve better performance with fewer heads. Compared to other two-tower structure models, its computation cost is comparable and it does not incur additional storage cost.

\subsection{Model Study (RQ2)}
To further understand the performance of FIT, we perform extensive ablation experiments to verify the contribution of each component. Additional analyses are also conducted on the key training techniques and model variables. 

\subsubsection{Component Effectiveness Analysis} 
In this section, ablation studies are shown to investigate the effects of each component of FIT. We compare FIT with the following variants:
 
\begin{itemize}[leftmargin=1em]
  \item 
    \begin{math}
        \mathbf{w/o\, MQM:}
    \end{math}
     remove MQM and transform the embedding vectors list of user sequence feature via a average pooling layer.
      \item 
    \begin{math}
        \mathbf{w/o\, LSS:}
    \end{math}
      remove LSS and the output layer is the same as the Two-Tower.
      \item 
    \begin{math}
        \mathbf{w\,SumMax:}
    \end{math}
     replace LSS with the sum-max similarity score.
      \item 
    \begin{math}
        \mathbf{w\,FC:}
    \end{math}
     replace LSS with a FC layer after the similarity matrix between the user and item representations. 
\end{itemize}

Performances with key components removed or replaced are shown in Figure \ref{fig:4}. The results indicate that FIT consistently performs better than all the variants on both datasets, demonstrating the effectiveness of the proposed components in FIT. After removing the MQM that achieves user-item early interaction and captures diverse signals effectively from historical sequence features, the model's performance significantly declines. Moreover, in comparison with other late interaction structures, FIT achieves the best result, proving that LSS is closer to the theoretical optimum.

%\begin{figure}[h]
%  \centering
%  \includegraphics[width=\linewidth]{ablation.png}
%  \caption{The albation study of FIT on Taobao Dataset}
%\end{figure}

% 图片横坐标改成training steps%
\subsubsection{Effectiveness Analysis of Training Techniques} To verify the effectiveness of the training techniques in MQM, we show the performance of the model on the validation dataset with different structures as the training steps increase.

\begin{figure}[t]
    \centering
    %%\setlength{\abovecaptionskip}{0.2cm}
    %%\setlength{\belowcaptionskip}{-0.5cm}
    %% \subfigure[Softmax Weighting with Temperature] {
    %%    \includegraphics[width=0.46\linewidth]{training_techniques_1.png}
    %%    \label{fig:5a}
    %% }
    %% \subfigure[Parameter-free Self-attention] {
    %%     \includegraphics[width=0.47\linewidth]{training_techniques_2.png}
    %%     \label{fig:5b}
    %% }
    \includegraphics[width=\linewidth]{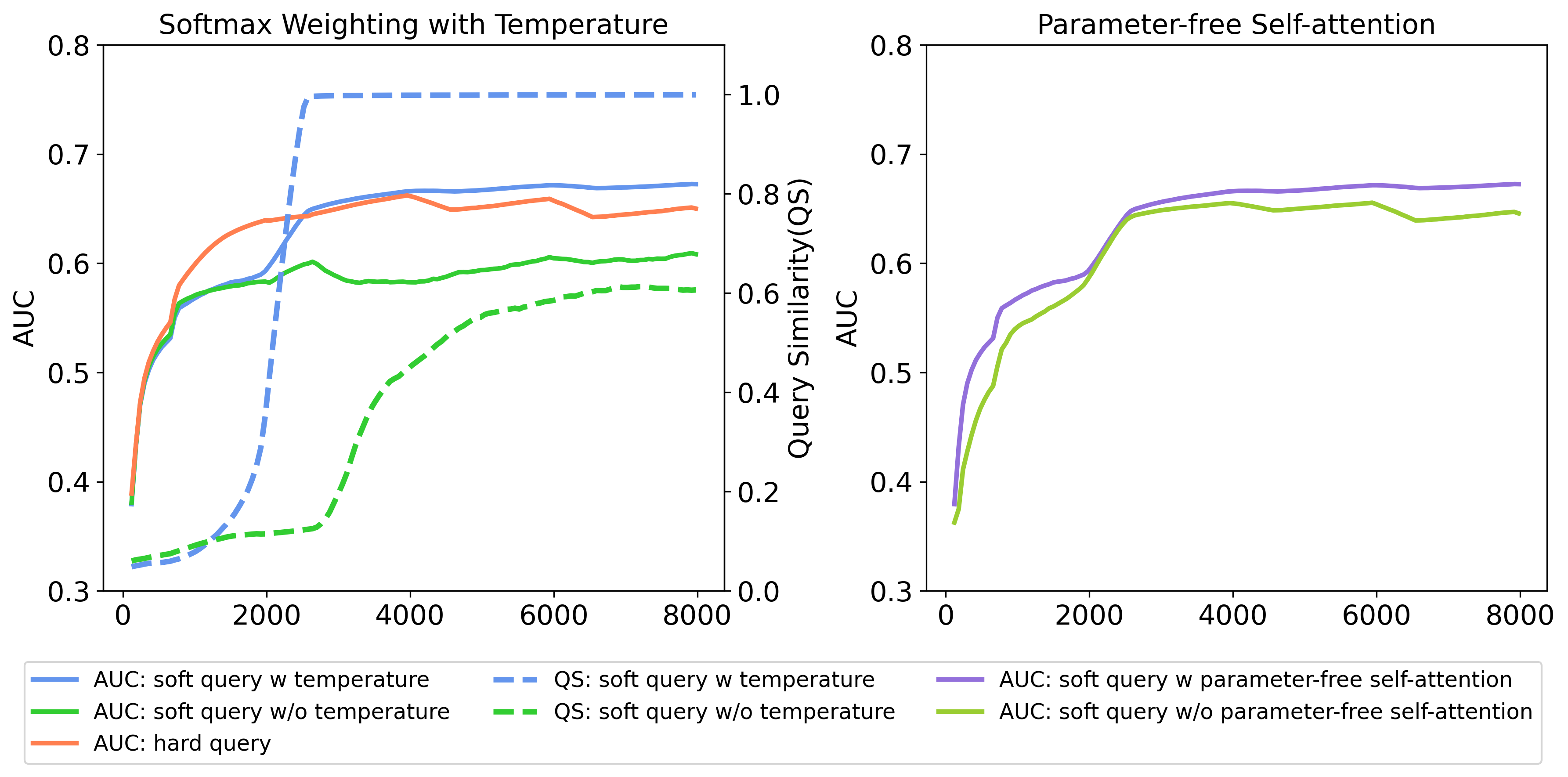}
    \caption{The effectiveness analysis of training techniques in FIT on Taobao.}
    \label{fig:5}
\end{figure}

The left part of Figure \ref{fig:5} shows that the model achieves the best result (AUC) when applying the soft query with decaying temperature, and the query similarity score (QS) between the soft query and the hard query reaches almost 1.0. The query similarity score is calculated as follows:
\begin{equation}
    QS = \frac{\mathbf{\tilde{q}^*}\cdot\mathbf{q}_i^*}{\|\mathbf{\tilde{q}^*}\| \|\mathbf{q}_i^*\|}.
\end{equation}
The greater the query similarity score, the higher the consistency between the training and inference phases. While directly using the hard query for training ensures complete consistency between training and inference, it does not yield optimal model performance because the item meta matrix cannot be fully trained as mentioned in Section \ref{sec:4.1}. Similarly, employing the soft query without temperature results in poor model performance due to the inconsistencies between training and inference.

%% \begin{figure}[t]
%%     \centering
%%     %%\setlength{\abovecaptionskip}{0.2cm}
%%     %%\setlength{\belowcaptionskip}{-0.5cm}
%%     \includegraphics[width=\linewidth]{training_techniques2.png}
%%     \caption{The effectiveness analysis of training techniques in FIT on Taobao.}
%%     \label{fig:5}
%% \end{figure}

In Figure \ref{fig:5}, the right part shows the performance of the model with and without parameter-free self-attention. We can see that the model with parameter-free self-attention converges faster at the beginning and finally obtains better performance.

\begin{figure}[h]
  \centering
  \includegraphics[width=\linewidth]{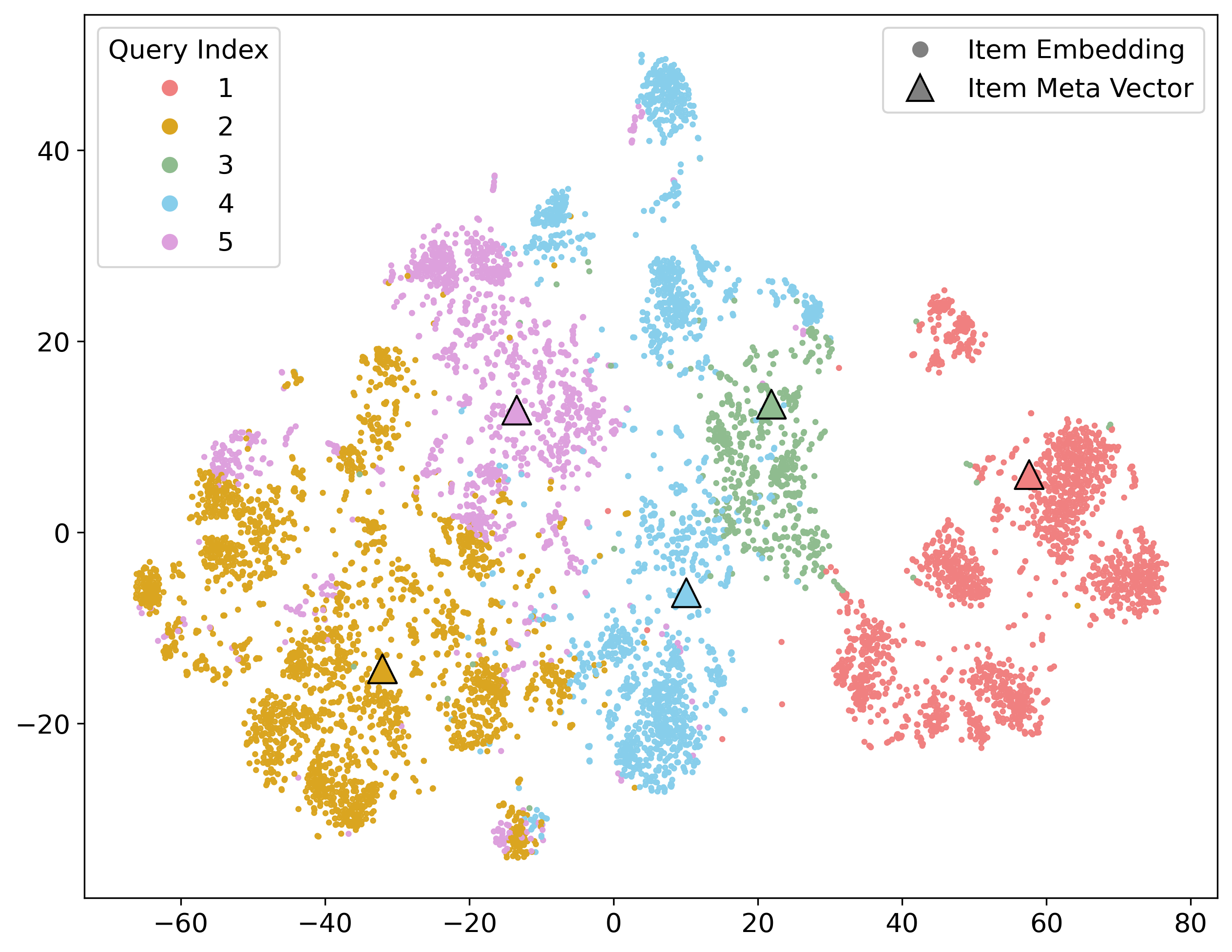}
  \caption{Visualization of the Item Meta Matrix}
  \label{fig:6}
\end{figure}
\subsubsection{Visualization of the Item Meta Matrix}
In order to have deep insights of FIT, we randomly select 50,000 items on the Taobao dataset and visualize their embedding vectors and the corresponding item meta vectors (hard queries) from the item meta matrix by projecting into a two-dimensional space with t-SNE\cite{van2008visualizing}. As illustrated in Figure \ref{fig:6}, the distribution of the item meta vectors is relatively uniform without all items collapsing into one group. The item embedding vectors mapped to the same query index are basically in the same cluster and the item meta vector is at the center of the item embedding vectors set. It demonstrates that the item meta matrix learns effective information from the item inputs and the two-tower model can complete expressive early interaction by using this meaningful representation of the clustered item set in the user tower to improve the performance.

\subsection{Hyper-parameter Study (RQ3)}
\label{sec:5.5}
To assess the impact of the key hyper-parameters on the performance of FIT, we conduct a series of experiments with different settings on the two largest datasets: Amazon Books and Taobao. The size of the item meta vectors in MQM and the number of heads in LSS are the two most important hyper-parameters of FIT. 
 
\subsubsection{Size of Item Meta Vectors in MQM} 
\begin{figure}[t]
    \centering
    %%\setlength{\abovecaptionskip}{0.2cm}
    %%\setlength{\belowcaptionskip}{-0.5cm}
    % \subfigure[Amazon Books] {
    %     \includegraphics[width=0.45\linewidth]{taobao_cluster_dimisions.png}
    %     \label{fig:7a}
    % }
    % \subfigure[Taobao] {
    %     \includegraphics[width=0.45\linewidth]{taobao_cluster_dimisions.png}
    %     \label{fig:7b}
    % }
    \centering
  \includegraphics[width=\linewidth]{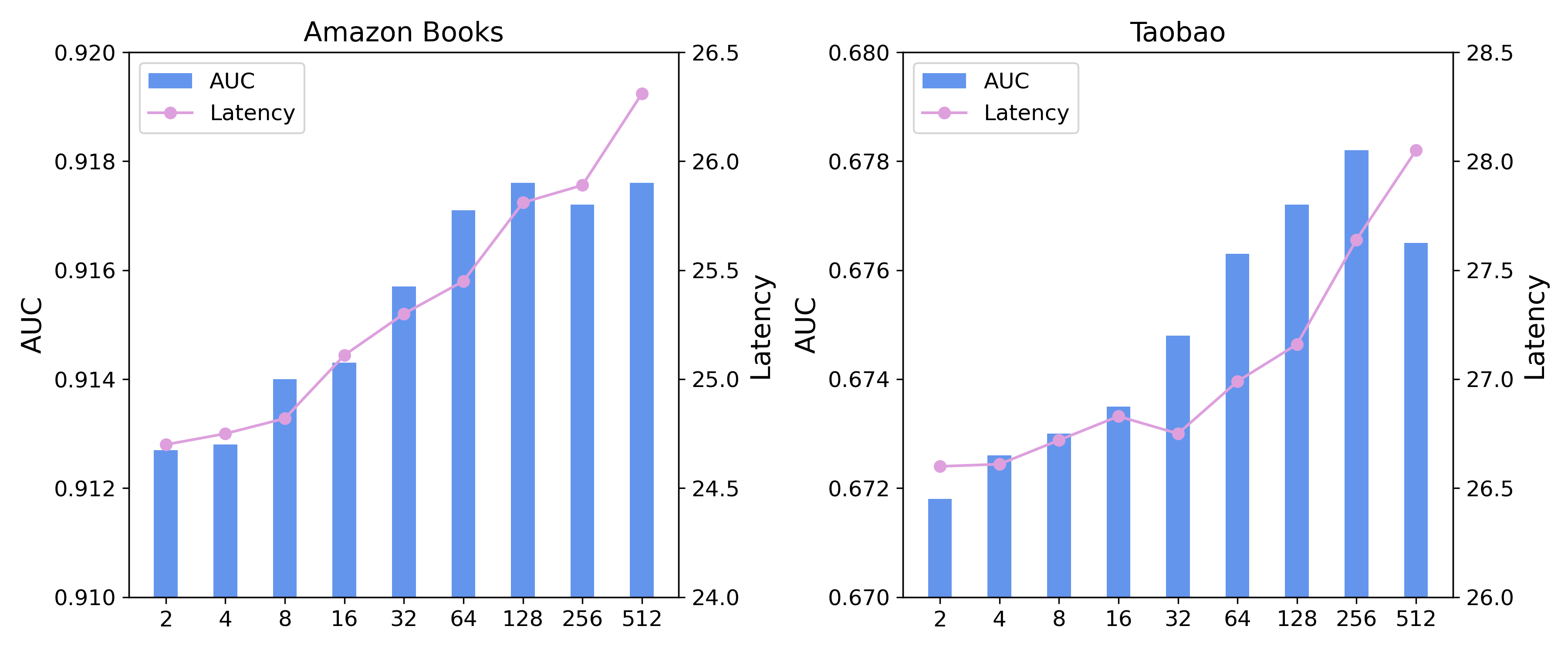}
 
    \caption{Performance and inference latency comparison with increased item meta vector size on Amazon Books and Taobao Dataset}
    \label{fig:7}
\end{figure}

The item meta matrix captures generalized information from the item tower, while an item meta vector represents the aggregated representation of an item cluster. The size of item meta vectors which denotes the item cluster number affects both the model performance and the inference efficiency. When the item meta vector size is larger, the granularity of feature interactions becomes finer, enhancing the expressive capability. However, it also increases the risk of overfitting, and excessively large size can result in additional inference costs. As shown in Figure \ref{fig:7}, as the vector size increases, the model performance gradually becomes better and then declines. The optimal performance can be achieved when the vector size reaches 256. The models with larger vector size tend to overfit and the inference latency can considerably increase. It is noteworthy that we report the model performance of FIT with item meta vector size set to 64 in Section \ref{sec:5.2}. This is not the optimal hyper-parameter setting in terms of performance, but it maintains fairly good inference efficiency. When the requirements for inference latency in online systems are not so strict, the item meta vector size can be increased to achieve better performance.

\subsubsection{Number of Heads in LSS}
\begin{figure}[t]
    \centering
    \includegraphics[width=\linewidth]{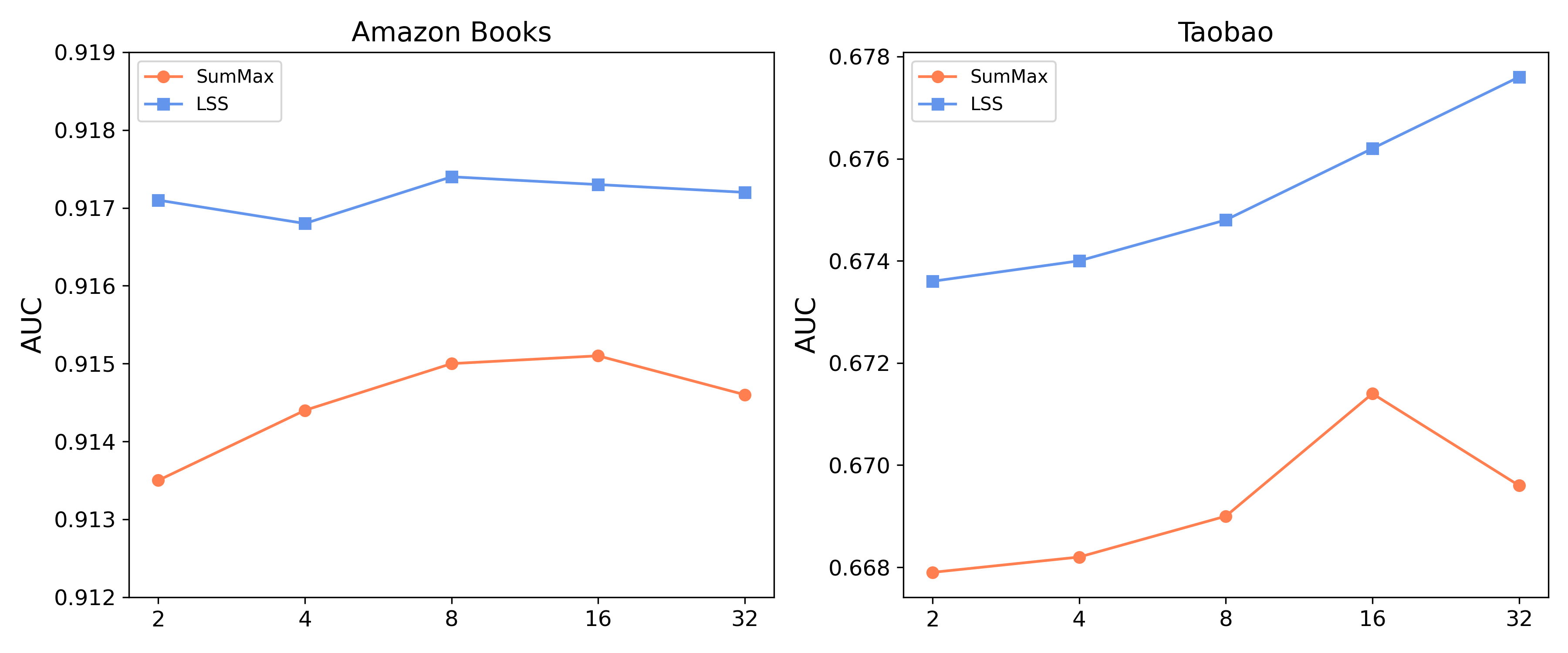}
    \caption{Performance comparison of different late-interaction structures with increased head size}
    \label{fig:8}
\end{figure}

The number of heads mainly influences the model performance and storage cost of the item representations. As the number increases, the model is capable of obtaining more diverse information and achieves better results. However, this comes with the trade-off of higher storage costs and an elevated risk of overfitting. We investigate the impact of an increased number of heads on different late interaction structures, as illustrated in Figure \ref{fig:8}. It can be observed that LSS significantly outperforms the sum-max similarity score with all head size settings. Moreover, with an increased number of heads, models can achieve better performance but are also more prone to overfitting and LSS is more likely to overfit than the sum-max similarity score, especially on the Amazon Books dataset, which has fewer data. Notably, we present the model performance of FIT with head size set to 2 in Section \ref{sec:5.2}. It is not the best hyper-parameter for performance, yet it can keep the storage cost equivalent to the two-tower model. When storage space is sufficient, FIT with more heads can yield improved results.

% \usepackage{multirow}
% \begin{table*} 
%   \caption{zheli tian yidui dongxi xxxxx}
%   \label{tab:commands}
%   \begin{tabular}{cccc}
%     \hline
%      \multirow{2}{*}{Model} & \multicolumn{3}{c|}{MovieLens} \\ \cline(2-4)
%      & AUC  & logloss & RelaImpr \\ \hline 
%      DSSM & 0.5 & 0.5 & 0 \\ \hline 
%   \end{tabular}
% \end{table*}

% \begin{figure}[t]
%   \centering
%   \includegraphics[width=\linewidth]{heads}
%   \caption{taobao heads\_picture}
% \end{figure}

%\begin{figure}[t]
%  \centering
%  \includegraphics[width=\linewidth]{temperature5}
%  \caption{temperature\_picture}
%  
%\end{figure}
%
%\begin{figure}[t]
%  \centering
%  \includegraphics[width=\linewidth]{temperature5.png}
%  \caption{self\_attention}
%  
%\end{figure}

\section{CONCLUSION}
In this paper, we propose a novel architecture named learnable Fully Interacted Two-tower Model (FIT) for large-scale pre-ranking recommendation systems. The classic two-tower model utilizes the user-item decoupling structure to ensure efficiency, but the model's performance is limited due to a lack of information interaction between the two towers. Our proposed FIT significantly enhances the interaction signals between the two towers through expressive early interaction and effective late interaction, improving model performance while maintaining high efficiency. Specifically, we design a Meta Query Module (MQM) to make arbitrary interaction methods available while maintaining user-item decoupling architecture, which considerably enhances the early interactions
between the two towers. In addition, we introduce a Lightweight Similarity Scorer (LSS) to effectively capture the late interaction signals to further improve the model performance. Finally, extensive experiments on several public datasets demonstrate the superiority of our proposed FIT over state-of-the-art methods. Further analysis also validates the effectiveness of the core designed techniques.

%%
%% The next two lines define the bibliography style to be used, and
%% the bibliography file.
\bibliographystyle{ACM-Reference-Format}
\balance
\bibliography{sample-base}

%%
%% If your work has an appendix, this is the place to put it.
\end{document}